\RequirePackage{fix-cm}
\documentclass[smallextended]{svjour3}       
\smartqed  

\usepackage[normalem]{ulem} 
\usepackage{graphicx}
\usepackage{multirow}
\usepackage{todonotes}
\usepackage{tcolorbox}
\usepackage{hyperref}

\usepackage[resetlabels,labeled]{multibib}
\newcites{New}{Reviewed Papers}

\begin{document}

\title{Modelling Guidance in Software Engineering: A Systematic Literature Review}

\author{Shalini Chakraborty \and Grischa Liebel}
\date{September 2019}

\institute{S. Chakraborty \at
              Reykjavik University\\ Menntavegur 1, 102 Reykjav\'{i}k, Iceland\\
              \email{shalini19@ru.is}
\and 
G. Liebel \at
              Reykjavik University\\ Menntavegur 1, 102 Reykjav\'{i}k, Iceland \\ ORCID: 0000-0002-3884-815X \\
              \email{grischal@ru.is}      
}

\date{Received: date / Accepted: date}

\maketitle
\begin{abstract}
Despite potential benefits in Software Engineering (SE), adoption of software modelling in industry is low. Technical issues such as tool support have gained significant research before, but individual guidance and training have received little attention. 
As a first step towards providing the necessary guidance in modelling, we conduct a systematic literature review (SLR) to explore the current state of the art. We searched academic literature for modelling guidance, and selected 25 papers for full-text screenin through three rounds of selection. We find research on modelling guidance to be fragmented, with inconsistent usage of terminology, and a lack of empirical validation or supporting evidence. 
We outline the different dimensions commonly used to provide guidance on software modelling.
Additionally, we provide definitions of the three terms modelling \emph{method}, \emph{style}, and \emph{guideline} as current literature lacks a well-defined distinction between them.
These definitions can help distinguishing between important concepts and provide precise modelling guidance. 
\end{abstract}
\section{Introduction}
Despite the potential benefits of using models in Software Engineering (SE), adoption has been low, typically pointing to issues such as tool support, organisational resistance, and a lack of guidance/training \cite{hutchinson11a,hutchinson11b,whittle13,hutchinson14,mohagheghi13a,mohagheghi08a,liebel18sosym,liebel18survey}. 
Technical issues have historically received substantial attention in the modelling community, seen for instance by the large amount of tools for modelling and Model-Based Engineering (MBE).
However, work on guiding individuals in creating models receives typically only marginal attention.
For instance, Sch\"{a}tz et al. state that ``methodical guidelines are missing how to use suitable abstractions of (parts of) a cyber-physical systems at varying level of detail to enable the engineering of those systems with a sufficient level of confidence concerning the quality of the implemented systems''~\cite{schatz15}.
%

%
%

To better understand shortcomings in literature on guidance and training in modelling, this paper reports on a Systematic Literature Review (SLR) \cite{kitchenham2007guidelines} surveying work on guidelines, styles, or training approaches on creating models in SE.
We address the following research question \textbf{RQ: What kind of guidance exists in SE literature on model creation?}
%
To answer the question, we collected a total of 6109 papers starting from 1998 when the first UML conference was held.
We complemented our database search with snowballing~\cite{Wohlin2014guidelines} in the area of Business Process Modelling.
%

After several rounds of exclusion, we analysed 23 papers, finding that systematic guidance to create models is absent in SE literature.
Existing work proposes guidance for specific domains or problems, but generally lacks validation and/or empirical evidence.
In BPM, which partially overlaps with SE research, there exists initial empirical work investigating modelling styles and how cognitive processes affect model creation.
We further find that terminology is used inconsistently in the literature, with terms such as method, guidelines, or style being used inconsistently and seemingly arbitrary.
To address this, we define \emph{modelling method}, \emph{modelling guideline}, and \emph{modelling style}.

Our SLR shows that further work is necessary in several directions in the modelling community.
First, we see that work on modelling needs to be conducted more systematically.
Our definitions presented in this paper can help supporting reaching a common terminology.
Secondly, empirical evidence and validation is needed.
Currently, we find mainly solution proposals that are at best substantiated through simplified examples.

The rest of the paper is structured as follows. In Section~\ref{sec:Related_Work} we discuss the lack of guidance in software modelling through related work. Section~\ref{sec:Method} explains the methodology of our conducted SLR and validity threats. We present the results in Section~\ref{sec:Result}, followed by an overall discussion of the findings in Section~\ref{sec:Disc}. Finally, we conclude the paper in Section~\ref{sec:Concl} with future plans to use the findings suitably in software modelling. 
\section{Related Work}
\label{sec:Related_Work}
Text books on UML and object-oriented design commonly refer to heuristics for creating UML diagrams, e.g., \cite{rumbaugh91,bruegge09}.
For instance, Abbott's \cite{abbott83} strategy for identifying system objects by scanning informal descriptions for nouns is typically referred to.
These and similar heuristics are common for some diagram types, e.g., class diagrams, but often lacking for behaviour.
Additionally, there is, to our knowledge, no empirical evidence that these kind of heuristics support system analysis or other modelling tasks.

Previous work has investigated how to design modelling languages in a \emph{good} way, by studying both the notations themselves, e.g., \cite{moody09,storrle13,santos18,santos18b}, and how models or diagrams are read and understood by human subjects, e.g., \cite{maier14,storrle18,storrle14,storrle16,lohmeyer15}.
Moody \cite{moody09} defines in his Physics of Notation a number of principles for cognitively effective notations.
However, the value of the Physics of Notation is unclear \cite{storrle13,santos18,santos18b}.
For diagram understanding, St\"{o}rrle et al. \cite{maier14,storrle18,storrle14,storrle16} find that the size and layout quality of UML diagrams directly impacts understanding.
Outside of SE, Lohmeyer and Meboldt \cite{lohmeyer15} find that there are different patterns individuals follow when reading engineering drawings.
Understanding how models are read and understood can help to derive heuristics for the quality of a model.
However, this knowledge might not help guiding individuals to create \emph{good} models.
That is, while the desired outcome might be known, the process to arrive at this outcome remains unclear.

A way to describe how to create models are to provide reusable parts of a solution, or patterns, for different kinds of models.
In the area of patterns, there has been substantial research during the late 1990s, e.g., \cite{gamma95,douglass99,bordeleau99,dwyer1998property,dwyer1999patterns}.
The well-known Gang-of-Four design patterns \cite{gamma95} provide reusable patterns that should help to improve the design of object-oriented systems.
Douglass describes patterns for real-time behaviour diagrams~\cite{douglass99}, and Bordeleau~\cite{bordeleau99} proposes patterns that should aid a designer to move from a scenario-based specification to a state-based specification.
Dwyer et al.~\cite{dwyer1998property,dwyer1999patterns} propose property specification patterns, reusable logic constructs that can be re-used to specify properties in specifications, specifically the order and the occurrence of system events.
The work on patterns has evolved substantially since then, especially in the area of real-time systems, e.g., \cite{autili15,grunske2008specification,maoz15}.
However, patterns describe only a part of model creation, namely \emph{what} should be part of the model.
There are several other aspects that could be of interest during model creation, e.g., in which order to create parts of the model, what to include and what to exclude, how to decide on a reasonable level of abstraction, and individual differences that affect this process.

Towards understanding the differences between individuals when creating models, Pinggera et al.~\cite{Pinggera2013Styles,pinggera12,claes15} investigate BPM model creation.
The authors present an exploratory study with 115 students exploring how students create BPM.
They find distinct styles of modelling by performing cluster analysis on phases of model comprehension (where a modeller builds a mental model out of the domain behaviour), modelling (where a modeller maps the mental model to actual modelling constructs), and finally reconciliation (where a modeller acknowledges the process model).
\section{Research Methodology}
\label{sec:Method}
In this paper, we aim to answer the following research question:
\begin{center}
    \textbf{RQ: What kind of guidance exists in SE literature on model creation?}
\end{center}
To do so, we conduct an SLR, a secondary form of study used to identify and evaluate available research relevant to a certain research question or topic of interest \cite{kitchenham2007guidelines}.
SLRs are useful to obtain a detailed picture of a research area, and to integrate existing evidence \cite{Kuhrmann2017OnThePragmatic}.
They are increasingly common in SE~\cite{Kitchenham2009literature}, and specifically in the modelling community in the last decade, e.g., \cite{somogyi2020systematic,giraldo2014analysing,nguyen2015extensive,nguyen2013systematic,loniewski2010systematic}.
We chose an SLR over a mapping study, as we were interested in the detailed picture, not the broad coverage of a research area a mapping study provides \cite{Petersen2008Mapping}.

\begin{figure}[!ht]
\includegraphics[width=1.0\textwidth]{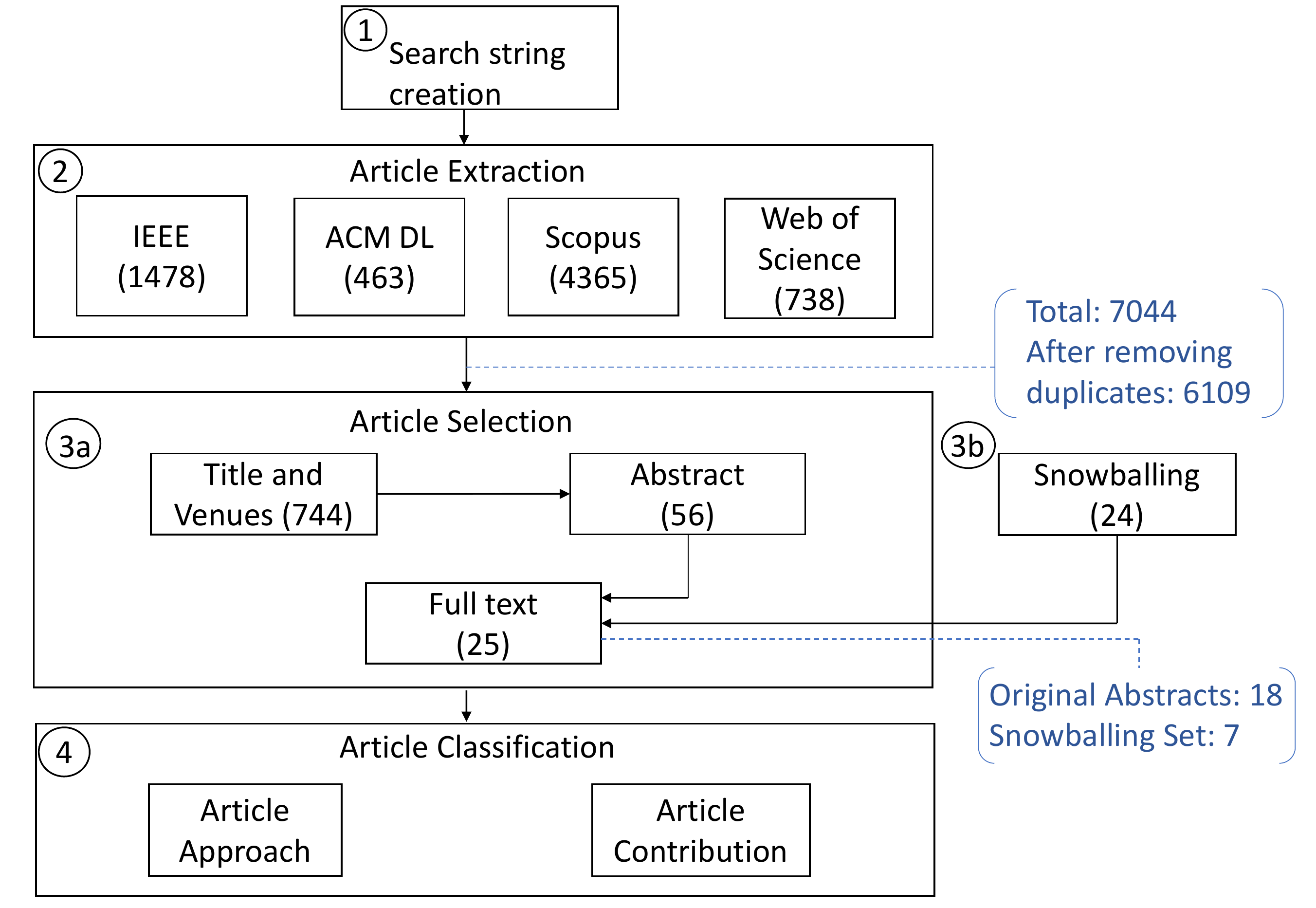}
\caption{Review Process}
\label{fig:Review process}
\end{figure}

We followed the steps proposed by Kitchenham and Charters \cite{kitchenham2007guidelines} to perform our SLR.
An overview of this process is depicted in Figure~\ref{fig:Review process}. 

In the following, we discuss the review steps in detail, starting with preliminary search.
\subsection{Preliminary search}
\label{sec:preliminary}
We started the review with preliminary research steps, by manually reading through the last five years\footnote{At the time of starting the review.} (2015 to 2019) of papers appearing in \emph{MODELS conference} (\url{http://modelsconference.org/}) and \emph{SoSyM journal} (\url{https://link.springer.com/journal/volumesAndIssues/10270}), the two prime venues for SE modelling research. 
We observed that papers use the terms \emph{framework}, \emph{guidelines}, \emph{method}, \emph{approach}, and \emph{pattern}.
%
Hence, we used these terms in our initial search string.
We searched within modelling research in SE, thus resulting in the following initial search string:\newline
\textbf{(``modeling''  OR  ``modelling''  OR  ``model-driven''  OR  ``model-based'')\newline  AND  (``framework''  OR  ``guidelines''  OR  ``method''  OR  ``approach'' OR ``pattern'')\newline  AND  (``software engineering'')}

Applying the initial search string on Scopus, IEEE Xplore, ACM Digital Library, and Web of Science resulted in 17713 papers, a result we deemed infeasible to analyse.
Therefore, we decided to make a number of adaptations to the search string.
First, we added three terms we had missed initially: \emph{creation}, \emph{training}, and \emph{styles}.
Adding creation and training is directly motivated by our research question and goal, while adding styles is motivated by existing work in BPM, i.e., by Pingerra et al.~\cite{Pinggera2013Styles}.
Secondly, to reduce the amount of found papers, we decided to remove \emph{method}, \emph{approach}, and \emph{pattern}.
The former two terms are relevant to our search, but their use is heavily conflated in SE research.
This can be seen as a sort of sampling from the population of papers that potentially discuss advice on model creation.
The third term, \emph{pattern}, is used heavily due to the popularity of design and architectural patterns.
We decided to exclude this line of research from our SLR, as it presents ``proven solutions'', and does not give direct guidance or advice for creating models.

\subsection{Search and Extraction}

%
%
%

%

The final search string we used is as follows:\newline
\textbf{(``modeling'' OR ``modelling'' OR ``model-driven'' OR ``model-based'')\newline AND (``guidelines'' OR ``training'' OR ``styles'' OR ``creation'')\newline AND (``software engineering'')}.
We adapted the search string to the four search engines we used, further limiting the search to English papers published from 1998, as the first UML conference was held in that year \cite{Jean1998uml}.
We followed the UK spelling of modelling throughout our review, but we used both UK and US spelling in our search string. Additionally, we added \textit{model-driven} and \textit{model-based} to target papers specifically aimed towards model-driven and model-based engineering (MDE \& MBE) as well.

We searched in title, abstract, and keywords for Scopus, IEEE Xplore, and ACM Digital Library, and in all fields offered by Web of Science.
For ACM Digital Library, we had to further exclude the keyword \emph{creation}, since one of the standardised ACM keywords (``CCS concepts'') is coined ``Software creation and its management''.
As it was not possible to exclude the CCS concepts from the search, we would get all papers that used this keyword prior to removing \emph{creation} from the search.
The overall search yielded a total of 7044 papers, leaving us with 6109 papers after removing duplicates.

\subsection{Paper Selection}
\label{sec: paper_selection}

%
We performed paper selection in three rounds with different inclusion and exclusion criteria for each round.
In the beginning of each round, both authors went through a random 5 per cent of the remaining papers to determine inter-rater agreement on the respective inclusion/exclusion criteria. 
We used Fleiss kappa to measure our agreement, a statistical measure used to evaluate the reliability of an agreement between a fixed number of raters \cite{kilicc2015kappa}.
We decided on a threshold value of $\kappa>0.7$ as a minimum agreement to continue the selection process.
This is a compromise between Fleiss~\cite{fleiss2013statistical} suggestion to understand kappa values of $>0.75$ as excellent, and suggestions of Landis and Koch~\cite{landis1977measurement} to label agreements between $0.61-0.80$ as substantial, both of which have been criticised for being arbitrary~\cite{gwet2010handbook}.
In case of lower values, we would discuss our disagreements, potentially refine the criteria, and then re-run the process with another random selection of papers.

In the first round, we excluded papers based on title and the paper venue.
We needed three rounds to reach a Fleiss Kappa of $\kappa \approx 0.73$.
Ultimately, we excluded papers in the first round if they matched any of the following criteria.
Note that we applied the criteria extremely conservative at this stage, to not exclude relevant papers.
\begin{enumerate}
        \item It is clear from the title that modelling/diagrams are not a contribution.
        \item The conference/journal is from a different field of science.
        \item The paper is not peer reviewed.
        \item The paper discusses properties of modelling languages or compares languages.
        \item The paper has a prose title, or clearly points into a different direction than investigated in our review.
        \item The paper is about descriptive models of a scientific or real-life concept, as opposed to a system. For instance, process improvement models such as CMMI, or models of course curricula.
        \item Modelling in a different domain of computer science, e.g., database models, machine learning models, or neural network models.
        \item Extended abstracts for tool demos or posters.
        \item Meta studies such as literature reviews, or comparative studies such as controlled experiments.
    \end{enumerate}
        %
        %
        %
        %
After applying these criteria, we were left with 744 papers for the second round of selection. 
    
In the second round, we considered the paper abstracts.
After two rounds of discussions and adapting the exclusion criteria, %
%
%
we reached a kappa of $\kappa \approx 0.79$.

The exclusion criteria for excluding based on abstracts are as follows.
    \begin{enumerate}
    \item Creation of models is not clearly mentioned in the abstract.
    \item It is clear from the abstract that the paper focuses on any of the following.
    \begin{enumerate}
        \item Language design/meta modelling
        \item Model transformations
        \item Secondary studies such as SLRs
        \item Controlled experiments or other comparative studies
        \item Descriptive models of a scientific or real-life concept, as opposed to a software system
        \item Architectural styles or design patterns
    \end{enumerate}
   \end{enumerate}
We excluded language design/meta modelling and model transformations, as we consider them special cases of modelling aimed at the meta modelling level.
Applying these criteria, we excluded 688 papers from the remaining 744 papers, leaving us with 56 papers for full-text reading.

After the second exclusion round, we found three papers from the area of BPM that fit into the scope of our review.
However, they were not included through the search, as they did not fit the software engineering keyword.
Nevertheless, to be able to compare SE-specific research with promising work in BPM (but outside SE), we took those three papers as a starting set for backwards and forwards snowballing \cite{Wohlin2014guidelines}.
That is, we searched both the references and citations of the three papers, including papers that fit our scope.
We used the same exclusion criteria as for the second round resulting in a set of 24 papers from BPM. 

In the third and last round, we extracted and read the full-text of the 56 remaining papers, as well as the 24 papers from the snowball search in BPM.
We used the same exclusion criteria as for the second round. We applied the criteria to each paper's full text. Furthermore, we excluded one paper, as an extended version of that paper was already included in the review.

This resulted in a final set of 18 papers from the original paper set, and 7 papers from the snowball search left for analysis.

    

%
\subsection{Data Extraction}
\label{sec:Extraction}
The final 25 papers we analysed are listed in Tables~\ref{tab:finalPaperSetOriginal} and \ref{tab:finalPaperSetSnowball}, in Appendix~\ref{app:papers}.
We provide the citations of all 25 papers in Appendix~\ref{app:papers}.
Both authors analysed all 25 papers, then discussed their disagreements.
We did not calculate inter-rater agreement at this step.
We deemed this unnecessary, as re-running the analysis several times would likely result in the authors memorising the discussions, not an actual assessment of inter-rater agreement.

%
For analysis, we classified each paper based on their \textbf{research approaches} and \textbf{research contributions}.
%
%
%
We use the classification of research approaches by Wieringa et al.~\cite{Wieringa2006classification} and later updated by Petersen et al.~\cite{PETERSEN2015MappingUpdate} to describe the contribution of the papers and thus the maturity of the research area.
Table~\ref{tab:paper_types_approach} summarises the different paper categories.

\begin{table}[ht!]
    \centering
    \begin{tabular}{|p{.25\linewidth }|p{.75\linewidth }|}
    \hline
        Research Approach & Definition\\
    \hline \hline
    Evaluation Research & Investigation of a problem employing case study, controlled experiment, survey etc in the industrial context, implementing a technique in practice, and evaluating the implementation, i.e., showing the benefits and drawbacks of the implementation. \\
    \hline
    Solution Proposals & Proposed solution (either novel or extension of an existing technique) with a relevant argument or a good example to show the solution's benefits. No empirical evaluation. \\
    \hline
    Validation Research & Novel research techniques used for experiments which are not established in practice (for example, studies conducted with students); investigation of properties of a solution proposal that has not been implemented yet.\\
    \hline
    Philosophical Papers & A conceptual framework that sketches a new way of looking at existing things.\\
    \hline
    Opinion Papers & An author's personal opinion whether a certain technique
    is good or bad, or how things should be done.\\
    \hline
    Experience Papers & An author's personal experience of how something has been done.\\
    \hline
    \end{tabular}
    \caption{Classification of Paper Types based on Wieringa et al.~\cite{Wieringa2006classification} and Petersen et al.~\cite{PETERSEN2015MappingUpdate}.}
    \label{tab:paper_types_approach}
\end{table}

\begin{figure}[!ht]
\centering
\includegraphics[width=.5\textwidth]{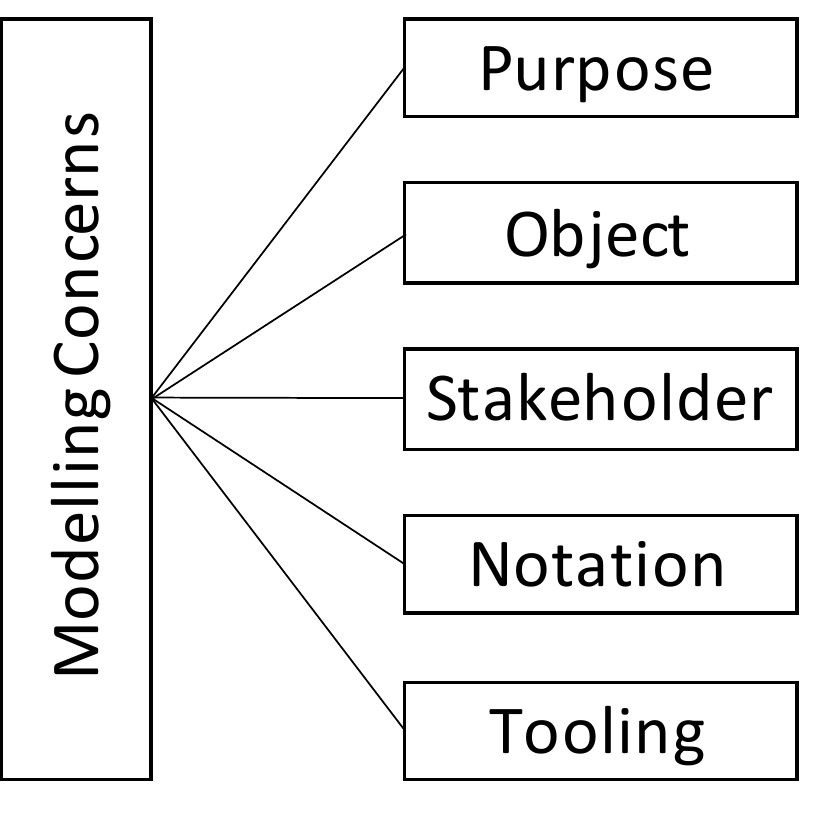}
\caption{Classification of Modelling Concerns. Adapted from \cite{liebel16lic}.}
\label{fig:contribution_classification}
\end{figure}

To provide a detailed view on what parts of modelling the papers contribute to, we break down the papers into different \emph{modelling concerns}. These concers are based on a classification we developed in previous work \cite{liebel16lic}. 
We do not claim that these concerns are exhaustive, but they serve as a useful tool to further break down a paper's contribution in terms of modelling.
The modelling concerns are depicted in a basic feature model notation in Figure~\ref{fig:contribution_classification}.
Solid lines from the \emph{modelling concerns} box depict optional features. That is, all of the features on the right-hand side of the figure are potential concerns of modelling that can be discussed in an paper.
The different concerns are as follows:
\begin{enumerate}
    \item Purpose: For what purpose is/are the model(s) created? For instance, models could serve as a software architecture description, or for communication within a development team. 
    \item Object: What is being described in the model? For instance, the model could describe an entire system, a subsystem or component, or a process.
    \item Stakeholder: Who is/are the primary stakeholders of the model? For example, a model could primarily serve as end-user documentation, or as a blueprint for developers.
    \item Notation: What modelling notation(s) is/are used? For instance, UML might be used.
    \item Tooling: Which tools are prescribed? For instance, Eclipse Papyrus might be prescribed due to a dependence on a custom plugin.
\end{enumerate}
%
%
To understand the modelling concerns we used open coding in the 25 papers.


%

%


\subsection{Threats to Validity}
\label{sec:Validity threats}
In this section, we describe the potential validity threats of our study and the steps we have taken to mitigate them. We follow the categorisation of threats into internal validity, construct validity, external validity, and reliability according to Runeson et al.~\cite{runeson2012case}. 

\subsubsection{Internal Validity}     
Internal validity reflects to what extend causal relationships are closely examined and other, unknown factors might impact the findings.

As a main means to reduce threats to internal validity, we tried to increase reliability of our exclusion steps by calculating inter-rater agreements and repeating the steps until a satisfactory level of agreement was reached between the two authors.
To avoid memory effects, we chose a new random sample to calculate inter-rater agreements after each round.

\subsubsection{Construct Validity}
Construct validity reflects to what extent the measures represent the construct investigated.

In this paper, we investigate model creation, guidance, and related topics.
A threat to the validity of our study is that these topics are not described using an established, standardised terminology in the modelling community.
Therefore, it might in some cases be difficult to decide when an paper in fact provides guidance for model creation, and when not.
To reduce this threat, we used open coding for data analysis, without relying on fixed keywords being used in the respective papers.

\subsubsection{External Validity} External validity is reflecting to what extent findings can be generalised beyond the concrete sample.

In case of our literature review, we use the major digital libraries for data collection, which should lead to a representative sample of publications in software engineering.
However, a potential threat to validity is our decision to exclude the conflated keywords ``approach'' and ``method''.
Our impression is that these keywords are used arbitrarily.
Hence, excluding them should not affect our results.
    
\subsubsection{Reliability}
Reliability describes the degree to which similar results would be obtained if the same study would be repeated, by the same or by other researchers.

The different steps of the data collection and analysis are clearly described in the previous subsections, so that we are confident that other researchers could repeat the study.
Nevertheless, there is subjectivity to several parts of our study, in particular the exclusion and analysis steps of the review.
We tried to be conservative in our exclusion, and calculated inter-rater agreement at all exclusion steps.
Furthermore, the analysis was performed by multiple authors, and disagreements discussed.
\section{Results}
\label{sec:Result}

%
In this section, we present the results of the SLR. First, we classify the final 25 papers in Section~\ref{sec:classification}. Based on the little empirical evidence we find in the modelling literature, we then introduce definitions for modelling guidelines, methods and styles in Section~\ref{sec:Definitions}. Finally, we answer the RQ in Section~\ref{sec:Ans_RQ}. 

\subsection{Paper Classification}
\label{sec:classification}
\begin{figure}[!ht]
\includegraphics[width=1.0\textwidth]{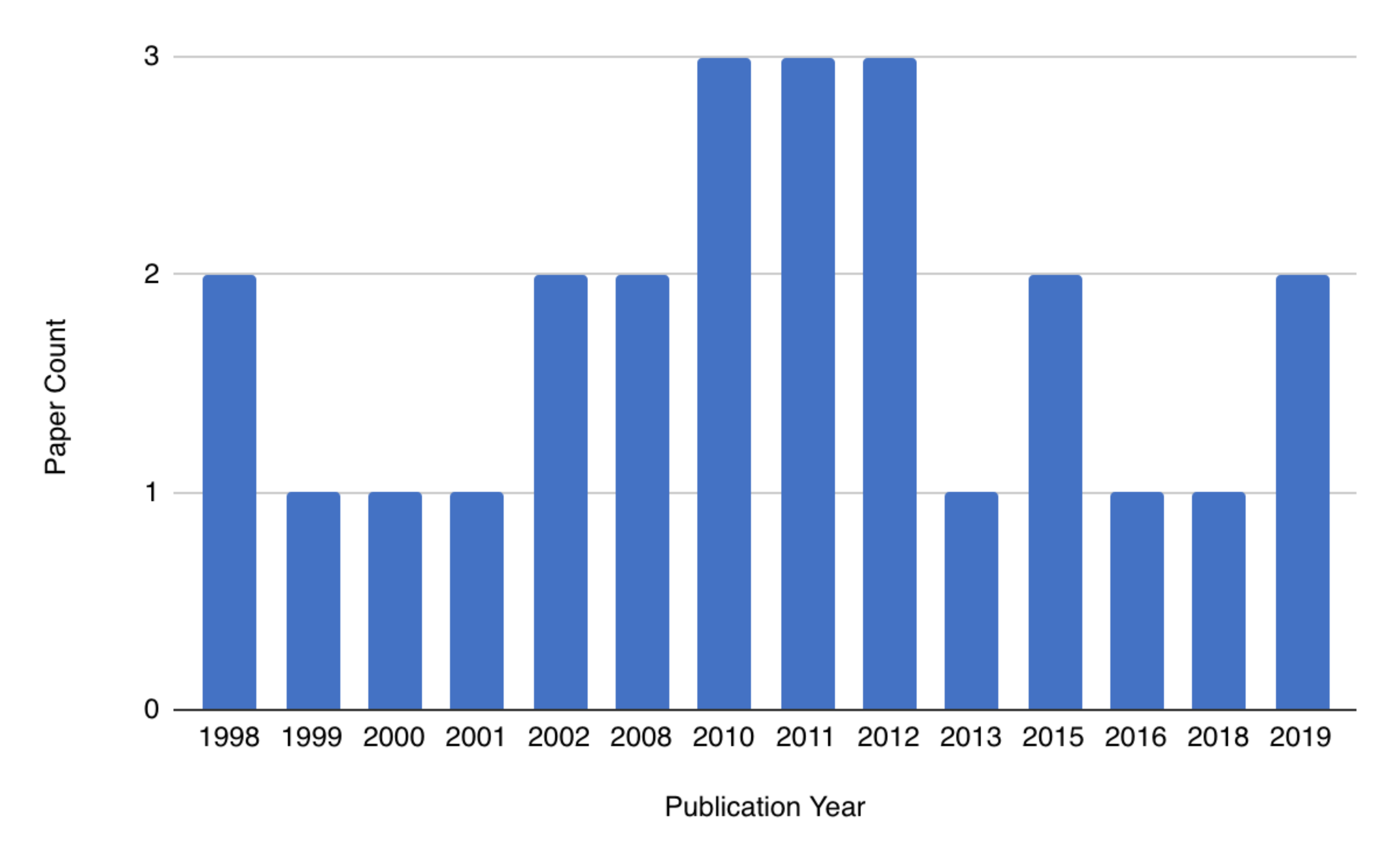}
\caption{Paper count per year. Years without papers are omitted.}
\label{fig:years}
\end{figure}

We ultimately analysed 18 papers out of initial 6109, with an additional 7 papers added through snowball search. 
The publication years of our final 25 papers range from 1998 to 2019, with the majority of the papers from 2008 to 2015.
Figure~\ref{fig:years} depicts the actual counts, where years without paper are omitted.

There is no clear pattern with respect to publication venues in the data.
Venues targeted towards modelling, such as the ER and UML/Models conference series or the Software and Systems Modeling journal are present, as well as general software engineering venues, such as ICSE or Transactions on Software Engineering.

Table~\ref{tab:paper_type_count} summarises the types of the 25 papers.
%
%
\begin{table}[ht!]
    \centering
    \begin{tabular}{|l|c|}
    \hline
        Paper Type & \# papers \\
    \hline \hline
    Evaluation Research (EV) &  7 \\
    \hline
    Solution Proposals (SP)&  14 \\
    \hline
    Validation Research (VR)&  1 \\
    \hline
    Philosophical Papers (PH)& 1 \\
    \hline
    Opinion Papers (OP)& 0 \\
    \hline
    Experience Papers (EX)& 2 \\
    \hline
    \end{tabular}
    \caption{Paper Types in our Analysis (based on Wieringa et al.~\cite{Wieringa2006classification}).}
    \label{tab:paper_type_count}
\end{table}
%

%
In terms of modelling concerns, we find that the model purpose, the modelling notation, and the object that is modelled are typically described in the papers, or can be derived from the descriptions.
In contrast, stakeholders and tooling are rarely reported.

The included papers state several \emph{purposes} for the models created.
Most commonly (9 out of 25 papers), there is no exact purpose stated for the constructed models, i.e., the purpose is to model parts of or a complete system.
The guidance provided in the papers thus aims broadly at improving the quality of the resulting models, or make the process more systematic.
The remaining papers state specific purposes for their constructed models.
These purposes vary widely, e.g., models for simulation, formal verification, requirements and system analysis, avoiding inconsistencies between models, and generate (parts of) an application.

Given our snowball search in the area of BPM, business processes are the dominating \emph{object} modelled in the included papers.
This is followed by six papers where modelling of embedded control systems is reported.
The remainder of the papers focuses on a mix of different models, such as use cases, architecture, and requirements.

Only one paper reports a specific \emph{stakeholder}.
Additionally, several papers refer to general terms such as \emph{engineers, programmers, domain experts, or software developers}.
That is, it is often unclear who creates, modifies, reads, or otherwise comes in contact with the proposed models.

9 of our 25 papers discuss models in UML \emph{notation}, or extensions thereof, such as UML profiles or the UML-RT extensions.
This is followed by 4 papers using BPMN.
Finally, several papers do not focus on a specific notation, or propose their own notations.

\emph{Tools} are rarely reported in our data set.
Only 5 out of 25 papers report a specific tool, 4 of which are specialised tools developed for the presented approaches.
Of the remaining 20 papers, most are tool agnostic, e.g., as they target general quality improvement to a specific type of model or as they do not rely on specific tool features.

\subsection{Definitions of Key Terms in Model Creation}
\label{sec:Definitions}

Starting from the initial keyword search and continuing throughout the entire analysis, we noticed that terms such as \emph{method}, \emph{guidelines}, and \emph{style} are used seemingly arbitrary in the community.
Therefore, we decided to provide definitions for a number of key terms, in order to allow us to better structure the literature, and to support future work in the area.

Modelling is a creative task, which requires thinking and problem solving, both of which can affect the quality of the model.
In the general literature on problem solving, Isaksen et al. \cite{Isaksen2016AnEO} discuss that, ``...when individuals, in both school and corporate settings, understand their own style of problem solving, they are able to learn and apply process tools more effectively, and when teams appreciate the styles of their individual members, their problem solving efforts are enhanced''. That is, styles in problem solving differ between individuals, which makes it necessary to distinguish from guidelines or methods, where specific steps or activities are prescribed on good practice.

We define the term \emph{modelling style} in reference to cognitive styles, which are ``an individual’s preferred way of gathering, processing and evaluating data''~\cite{Allison2006CognitiveStyle}.
Messik~\cite{Messick1984TheNO} defines cognitive styles as ``consistent individual differences in preferred ways of organizing and processing information and experience''.
Accordingly, we define modelling styles as follows.
\begin{tcolorbox}
\textbf{Modelling styles} are ``consistent individual differences in preferred ways of creating and processing software models''.
\end{tcolorbox}
Here, creation can include aspects such as the choice of modelling language, preferred elements of those languages, of tools, and organising the created model (i.e., layouting, or using a specific subset of modelling elements from a given language).
Processing includes reading, understanding and validating models.

In contrast, we define modelling methods as follows.
\begin{tcolorbox}
\textbf{Modelling methods} are ``sets of steps or constraints on how one or multiple models are created''.
\end{tcolorbox}
In contrast to a modelling style, a modelling method is identical across individuals, and does not depend on preferences.
However, modelling methods and modelling styles overlap and may conflict with each other, e.g., if an individual's preferred ways of creating a model are in conflict with the prescribed method.

Finally, for guidelines we use an existing dictionary definition. 
\begin{tcolorbox}
\textbf{Guidelines} are ``information intended to advise people on how something should be done or what something should be\footnote{\url{https://dictionary.cambridge.org/dictionary/english/guideline?q=Guidelines}}".
\end{tcolorbox}
Following this definition, modelling methods are always also guidelines, as they advise people how to create a model.
The opposite is not true, as guidelines do not necessarily have to be in the form of steps or constraints.

\subsection{Modelling Creation Guidance in Literature}
\label{sec:Ans_RQ}
After defining the terms modelling \emph{style}, \emph{method}, and \emph{guidelines} in Section~\ref{sec:Definitions}, we now re-visit the literature in terms of our RQ, i.e., \emph{What kind of guidance exists in SE literature on model creation?}.%

Based on the definitions, we found 10 papers containing guidelines (i.e., \cite{Fernandes2000ModelingEmbedded,Reggio2012Business,Guizzardi2011design,Schutte1998Guidelines,Das2018UMLRT,Zhao2019Design,Becker2003Guidelines,Firesmith1999Usecaseguidelines,Corallo2011Guidelines,Mendling20107pmg}), 12 papers describing modelling methods (i.e., \cite{Juhrisch2010Context,Machado2010Scenario,Hennicker2001UML,Fatwanto2008Arc,deng2019modeling,Goncalves2013Guide,Marincic2008Nonmonotonic,Wang2012Usecase,Wu2015Avionics,kaewkasi2002WWM,Milani2013Modelling,Rolland1999Goalmodeling}), and 1 paper investigating modelling styles (i.e., \cite{Pinggera2012Modelling}). The remaining 2 papers (i.e., \cite{Claes2015Structured,Soffer2011Towards}) discuss selected modelling concerns without providing any of the above.
%
In the following, we describe different concerns addressed by the included papers, and relate these to whether guidelines, methods, or styles are described.

\begin{table}[!ht]
\centering
    \begin{tabular}{ |p{2cm}|l|l|l|l|l|l|l|l|l| }
\hline
 \hline
  Paper & \cite{Juhrisch2010Context} & \cite{Machado2010Scenario} & \cite{Hennicker2001UML} & \cite{Fatwanto2008Arc} & \cite{deng2019modeling} & \cite{Fernandes2000ModelingEmbedded} & \cite{Reggio2012Business} & \cite{Guizzardi2011design} & \cite{Goncalves2013Guide} \\
 \hline
 Guidance Type & M & M & M & M & M & G & G & G & M\\
 \hline
 Paper Type & SP & SP & SP & SP & SP & EX & SP & SP & SP \\
 \hline
 \hline
 What vs. How & Both & What & Both & What & What & What & What & What & Both \\
 \hline
 Views \newline/Perspectives & No & No & Yes & Yes & Yes & No & No & No & No \\
 \hline
 Decomposition /Refinement & Yes & Yes & No & No & Yes & Yes & No & Yes & No \\
 \hline
Practices \newline/Anti-patterns & No & No & Yes & No & No & No & Yes & No & No \\
 \hline
 Cognition & No & No & No & No & No & No & No & No & No \\
 \hline
 \end{tabular}
 \caption{Paper Overview, Part 1. Guidance type consists of Methods (M), Guidelines (G), and Styles (S). Paper types are listed according to Table~\ref{tab:paper_type_count}.}
 \label{tab:PaperOverview1}
\end{table}

\begin{table}[!ht]
\centering
    \begin{tabular}{ |p{2cm}|l|l|l|l|l|l|l|l|l| }
\hline
Paper & \cite{Schutte1998Guidelines} & \cite{Marincic2008Nonmonotonic} & \cite{Das2018UMLRT} & \cite{Zhao2019Design} & \cite{Becker2003Guidelines} & \cite{Wang2012Usecase} & \cite{Wu2015Avionics} & \cite{Firesmith1999Usecaseguidelines} & \cite{kaewkasi2002WWM} \\
 \hline
 \hline
 Guidance Type & G & M & G & G & G & M & M & G & M \\
 \hline
 Paper Type & SP & SP & EV & EV & SP & SP & EV & EX & SP \\
 \hline
 \hline
 What vs. How & Both & What & How & How & What & What & Both & Both & What \\
 \hline
 Views \newline/Perspectives & No & No & No & No & No & Yes & No & No & No \\
 \hline
 Decomposition /Refinement& No & Yes & No & No & No & Yes & No & No & No\\
 \hline
Practices \newline/Anti-patterns & Yes & No & Yes & Yes & Yes & No & No & Yes & No \\
 \hline
 Cognition & No & No & No & No & No & No & No & No & No \\
 \hline
 \end{tabular}
 \caption{Paper Overview, Part 2. Guidance type consists of Methods (M), Guidelines (G), and Styles (S). Paper types are listed according to Table~\ref{tab:paper_type_count}.}
 \label{tab:PaperOverview2}
\end{table}

\begin{table}[!ht]
\centering
    \begin{tabular}{ |p{2cm}|l|l|l|l|l|l|l|}
\hline
  Paper & \cite{Milani2013Modelling} & \cite{Corallo2011Guidelines} & \cite{Pinggera2012Modelling} & \cite{Claes2015Structured} & \cite{Soffer2011Towards} & \cite{Rolland1999Goalmodeling} & \cite{Mendling20107pmg} \\
 \hline
 \hline
 Guidance Type & M & G & S & N/A & N/A & M & G \\
 \hline
 Paper Type & EV & SP & EV & PH & VR & EV & EV \\
 \hline
 \hline
 What vs. How & How & How & N/A & N/A & N/A & How & How \\
 \hline
 Views \newline/Perspectives & No & Yes & N/A & N/A & N/A & No & No \\
 \hline
 Decomposition /Refinement& Yes & Yes & N/A & N/A & N/A & Yes & No \\
 \hline
Practices \newline/Anti-patterns & No & No & N/A & N/A & N/A & No & Yes \\
 \hline
 Cognition & No & No & Yes & Yes & Yes & No & No \\
 \hline
 \end{tabular}
 \caption{Paper Overview, Part 3. Guidance type consists of Methods (M), Guidelines (G), and Styles (S). Paper types are listed according to Table~\ref{tab:paper_type_count}.}
 \label{tab:PaperOverview3}
\end{table}

\subsubsection{Inconsistent Terminology}
As discussed in Section~\ref{sec:Definitions}, terminology is used inconsistently across the literature.
For instance, \cite{Pinggera2012Modelling} describes modelling style in a similar way as we defined it above, while the ``styles'' described in \cite{Reggio2012Business} are different levels of formality used in business process models.
That is, they do not refer to individual differences between modellers.
Similarly, the described ``method'' in \cite{Reggio2012Business} is not a set of steps or constraints and therefore rather falls under our definition of guidelines.
Finally, we also find ``guidelines'' described in \cite{Goncalves2013Guide} that fit our description of a method.

\subsubsection{What instead of How}
There is one major distinction that can be made between our included papers.
While some papers give concrete advice on how a single model (or models of the same type) should be created, 10 papers provide guidance on using different models/diagrams towards an aim, such as modelling an entire system.
That is, the latter type of papers typically focuses more on \emph{what} to use, e.g., in terms of diagram types.
As timing and order are important elements in these descriptions, several of these papers are method papers (6 out of 10).

For example, in \cite{Juhrisch2010Context}, the authors describe a method to systematically model service-oriented systems in a series of steps.
The method aims at avoiding inconsistencies between models, and has a natural order of steps.
That is, constraints are imposed at language level before the actual system models are created.
In \cite{Fatwanto2008Arc}, a six-step method for modelling software architecture is presented, composed of selecting an architectural style/pattern, defining architectural structural elements and rules, specifying data structures, specifying structural units, specifying mechanisms to support state models and timers, and eventually build the architecture model.
Again, different models are used at different steps of the method, and the order of the steps is important.
\cite{Guizzardi2011design} describe guidelines for the use of OntoUML. These guidelines are essentially restrictions on what parts of the OntoUML notation to use.
Neither of these three papers provides detailed guidance on \emph{how} the individual models are created.

Finally, six papers mix guidance on what notations, diagrams, or formalisms to use with detailed instructions on how to use these.
For example, \cite{Schutte1998Guidelines} aims to improve the quality of information models by providing guidelines in the form of best practices.
These best practices include instructions \emph{what} diagrams to create, and \emph{how} to ensure several of them are consistent.
\cite{Firesmith1999Usecaseguidelines} also provides guidelines, including \emph{what} to use (e.g., in terms of tooling), and \emph{how} to approach modelling of use cases.

\subsubsection{View-Based or Perspective-Based Guidance}
Another common way to provide modelling guidance, found in 5 out of 25 papers, is to propose modelling according to multiple different views or system concerns, similar to architectural documentation that is typically presented according to multiple dimensions or in multiple views \cite{bass2003software}.

For example, guidelines for increasing the quality of process models are presented in \cite{Becker2003Guidelines}, using both perspectives and views.
The six perspectives of correctness, relevance, economic efficiency, clarity, comparability, and systematic design are considered, as well as several modelling views, such as a data view, organisational view, and control view.
Similarly, \cite{Hennicker2001UML} proposes a method to semi-automatically create hypermedia applications by modelling the application using different views, such as the navigation and presentation views.

\subsubsection{Decomposition-Driven or Refinement-Driven Guidance}

A common approach to deal with complexity in computer science is to decompose a problem into sub-problems, or to incrementally refine a solution.
These two approaches are also reflected in guidance on modelling.
That is, several papers propose methods and guidelines in which an abstract, incomplete or informal model is incrementally refined.

For example, the method proposed in \cite{Machado2010Scenario} suggests to incrementally formalise behaviour, starting from textual rules, which are then refined into graphical models, and finally formalised using colored Petri nets.
Similarly, Marincic et al.~\cite{Marincic2008Nonmonotonic} suggest a step-wise process in which requirements for embedded control systems are refined.
A method to model cyber-physical systems in terms of abstraction levels is presented in \cite{deng2019modeling}, where abstract concepts such as initial requirements are modelled first, and then increasingly described in more detail.
Ultimately, the method allows for simulation of the CPS' behaviour.
%
%
A method that prescribes step-wise decomposition of business processes is described in \cite{Pinggera2012Modelling}.
Between each decomposition step, it is decided which parts should be expressed together in a process model.
\cite{Milani2013Modelling} describes decomposition steps, each of which is accompanied by guiding questions.

\subsubsection{Best Practices and Anti-Patterns}
Best practices and anti-patterns are common ways to describe the state of practice and guide practitioners, e.g., in core computer science areas such as programming \cite{sutter2004c,long2013java} and software design \cite{brown1998antipatterns}.
This is also reflected in our data set, with 8 papers suggesting such best practices or anti-patterns.

For example, in \cite{Reggio2012Business}, five modelling styles are proposed ranging from the so-called ``ultra light style'', where UML activity diagrams are supported with free text, to the so-called ``precise operational style'', where UML and OCL are used according to their semantics. 
While the authors call these \emph{styles}, they are essentially best practices on which parts of the official UML notation should be used depending on the functional context (who, when, where, how and why will the model be used).
The styles are based on the authors' understanding, without any empirical data supporting them.

\subsubsection{Considering Cognition}
Finally, three papers in our dataset explicitly consider the modeller's cognitive processes, or argue that it should be taken into consideration.

In \cite{Soffer2011Towards}, the authors suggest that a mental model is created when a problem is conceptualised.
The mental model is the mapped to the actual solution model.
Assuming this process, the authors argue that improving the mental model could lead to an improved domain understanding.
\cite{Claes2015Structured} argues that cognition needs to be taken into account to explain shortcomings in BPM.
Finally, cognition is taken into account explicitly as a part of the theoretical framework in \cite{Das2018UMLRT}, where three styles for BPM are extracted based on an observational study.
The results describe a ``high-efficiency'' style, a ``good layout but less efficiency'' style, and a ``neither good layout nor very efficient'' style.
The authors describe task-specific and modeller-specific factors behind each style, especially how the above mentioned two factors influence specific aspects of each style. 
\section{Discussion}
\label{sec:Disc}
The objective of our SLR is to investigate how much guidance there exists in the literature for model creation in SE. 
We discuss these findings with respect to a number of observed themes below.

\subsection{Little empirical evaluation}
Among 25 included papers, 14 are solution proposals (see Table \ref{tab:paper_type_count}), and only 1 contains validation research. These statistics demonstrate a lack of empirical evaluation present in the literature. 
In contrast, the papers obtained through snowball sampling in BPM show a higher amount of evaluation and validation, with 4 out of 7 papers being evaluation research, and one paper being validation research.
That is, the area of BPM seems to be more mature when it comes to supporting model creation through empirical studies.
Similarly, the only paper explicitly considering cognition and the construction of mental models were in the area of BPM.

This lack of empirical studies is comparably common in the area of software modelling.
Zhang et al.~\cite{Zhang2018Empirical} find that empirical research methods within SE are mostly applied to software maintenance, quality and testing. While software models and methods gained empirical attention\footnote{66 papers published from 2013-2017 at EMSE and ESEM.} but none of them investigating modelling style nor model creation. 

\subsection{Lack of categorisation and organisation}
In addition to a lack of empirical evaluation, we observe that terminology differs substantially.
\textbf{Guidelines}, \textbf{methods}, \textbf{frameworks} and other terms are used in a seemingly arbitrary fashion.
We address this issue by proposing definitions that make use of existing ones as much as possible.
Additionally, we outlined dimensions and properties of existing modelling guidelines and methods.
Common ways to provide guidance in modelling are to describe what diagrams should be created, to prescribe views or perspectives that should be considered, to encourage a stepwise decomposition or refinement, and to suggest general best practices or anti-patterns.
With the exception of stepwise decomposition/refinement, these forms of guidance can either follow a dedicated set of steps or not.
For instance, modelling guidelines could simply outline a few UML diagrams that need to be created in arbitrary order, while a modelling method could prescribe the same diagrams, but in a stepwise fashion.

%

\subsection{Limited consideration of cognition}
In BPM, substantial work exists that explicitly discusses the role of cognitive processes and individual preferences in creating business process models.
In the SE modelling community, we do not find such work outside of the BPM community.
On the one hand, we believe this relates to the relative constrained nature of BPM as opposed to system modelling on a general level.
That is, notations, level of abstraction, and purpose vary considerably in software and systems modelling and are likely to affect the cognitive processes and modelling styles of individuals.
Nevertheless, we consider it relevant to explore this direction more in the future.
Existing work from the BPM community can be used as valuable guidance to design similar studies in software modelling as a whole.  
%
%

%
\section{Conclusion}
\label{sec:Concl}
We conducted an SLR on guidance in software modelling.
From initially 6109 papers, we extracted 25 papers for full-text reading. 

We find that terminologies are used arbitrary and inconsistently, and that little empirically validated modelling guidance exists.
Existing papers use different dimensions to provide guidance for modellers, often in the form of prescribing which diagrams/notations to use, which views/perspectives to model, how to decompose or refine a model of a system, or by providing best practices or anti-patterns.
In the BPM community, cognition has been explicitly considered as an important aspect of the modelling process.
However, our results show that we lack similar work outside the BPM community.
Finally, to address the lack of consistent terminology, we provide definitions for modelling \emph{guidelines}, \emph{methods}, and \emph{styles}.
We hope that our work will highlight the inconsistencies in modelling guidance. Considering the success of modelling exploration in the BPM community, we believe similar attention and detailing is needed for software models. Our definitions will help researchers in the field to precisely articulate their works. 
%



\bibliographystyle{spmpsci}      
\bibliography{refs}
\appendix

\section{Paper Categorisation Details}
\label{app:papers}
In the following, we list the papers we used during our analysis.
Table~\ref{tab:finalPaperSetOriginal} shows the papers extracted during the original search, while Table~\ref{tab:finalPaperSetSnowball} shows the papers from the snowball search.

\begin{table}[!ht]
    \centering
    \begin{tabular}{ |p{1cm}|p{7cm}|p{5cm}| }
\hline
Paper & Titles & Authors \\ [0.5ex] 
 \hline \hline
    \cite{Juhrisch2010Context} & Context-based modeling: Introducing a novel modeling approach &   M.  Juhrisch  and  G.  Dietz  \\
 
    \cite{Machado2010Scenario}  & Scenario-based modeling in industrial information systems & R. J. Machado, J.o M. Fernandes, J. P. Barros and L. Gomes \\
 
    \cite{Hennicker2001UML} & A UML-Based Methodology for Hypermedia Design & R. Hennicker and N. Koch \\
 
    \cite{Fatwanto2008Arc} & Architecture modeling for translative model-driven development & A. Fatwanto and C. Boughton \\
 
    \cite{deng2019modeling} & Modeling and simulation of CPS based on SysML and modelica(KG) & F. Deng, Y. Yan, F. Gao and Linbo Wu \\
 
    \cite{Fernandes2000ModelingEmbedded} & Modeling Industrial Embedded Systems with UML & J. M. Fernande, R. J. Machado and H. D. Santos \\
    
    \cite{Reggio2012Business} & Business Process Modelling: Five Styles and a Method to Choose the Most Suitable One & G. Reggio, M. Leotta, F. Ricca and E. Astesiano \\
 
    \cite{Guizzardi2011design} & Design patterns and inductive modeling rules to support the construction of ontologically well-founded conceptual models in OntoUML & G. Guizzardi, A. P. das Graças, and R. S. S. Guizzardi \\
 
     \cite{Goncalves2013Guide} & Guidelines for modelling reactive systems with coloured Petri nets & M. P. Gon\c{c}alves and J. M. Fernandes\\
 
    \cite{Schutte1998Guidelines} & The guidelines of modeling - An approach to enhance the quality in information models & R. Schuette and T. Rotthowe \\
 
    \cite{Marincic2008Nonmonotonic} & Non-Monotonic Modelling from Initial Requirements: A Proposal and Comparison with Monotonic Modelling Methods & J. Marincic, A. Mader, H. Wupper and R. Wieringa \\
    
   \cite{Das2018UMLRT} & Model development guidelines for UML-RT: conventions, patterns and antipatterns & T. K. Das and J. Dingel \\
 
    \cite{Zhao2019Design} & Design guidelines for feature model construction: Exploring the relationship between feature model structure and structural complexity & X. Zhao and J. Gray \\
 
    \cite{Becker2003Guidelines} & Guidelines of business process modeling & J. Becker, M. Rosemann and C. van Uthmann \\
    
    \cite{Wang2012Usecase} & A modeling approach for use-cases model in UML & Z. Wang \\
    
    \cite{Wu2015Avionics} & A modeling methodology to facilitate safety-oriented architecture design of industrial avionics software & J. Wu, T. Yue, S. Ali and H. Zhang \\
    
    \cite{Firesmith1999Usecaseguidelines} & Use case modeling guidelines & D. G. Firesmith \\
    
    \cite{kaewkasi2002WWM} & WWM: a practical methodology for Web application modeling & C. Kaewkasi and W. Rivepiboon \\
 \hline
 \end{tabular}
 \caption{Selected Papers from Original Set}
 \label{tab:finalPaperSetOriginal}
\end{table}

\begin{table}[!ht]
    \centering
    \begin{tabular}{|p{1cm}|p{7cm}|p{5cm}|}
    \hline
        Paper & Titles & Authors \\ [0.5ex] 
    \hline \hline
       \cite{Soffer2011Towards} & Towards Understanding the Process of Process Modeling: Theoretical and Empirical Considerations & P. Soffer, M. Kaner and Y. Wand \\
       
        \cite{Rolland1999Goalmodeling} & Guiding goal modelling with scenarios & C. Rolland, C. Souveyet and C. Ben Achour \\
        
        \cite{Milani2013Modelling} & Modelling Families of Business Process Variants: A Decomposition Driven Method & F. Milani, M. D., N. Ahmed and R. Matulevi\v{c}ius\\
        
        \cite{Corallo2011Guidelines} & Guidelines of a Unified Approach for Product and Business Process Modeling in Complex Enterprise & A. Corallo, P. De Paolis, M. Ippoliti, M. Lazoi, M. Scalvenzi and G. Secundo \\
        
         \cite{Pinggera2012Modelling} & Modeling Styles in Business Process Modeling & J. Pinggera, P. Soffer, S. Zugal, B. Weber, M. Weidlich, D. Fahland, H. A. Reijers and J. Mendling \\
        
        \cite{Claes2015Structured} & The Structured Process Modeling Theory (SPMT) a cognitive view on why and how modelers benefit from structuring the process of process modeling & J. Claes, I. Vanderfeesten, F. Gailly, P. Grefen and G. Poels \\ 
        
        \cite{Mendling20107pmg} & Seven Process Modeling Guidelines (7PMG) & J. Mendling, H. A. Reijers and W. M. P. van der Aalst \\
        \hline
    \end{tabular}
    \caption{Selected Papers from Snowballing Set}
    \label{tab:finalPaperSetSnowball}
\end{table}
%


\end{document}